\documentclass[12pt]{article}

\usepackage{hyperref}
\usepackage{amssymb}
\usepackage{graphicx}

\begin{document}

%%%%%%%%%%%%%%%%%%%%%%%%%%%%%%%%%%
\renewcommand{\abstractname}{\hfill}

%%%% *****************************************************************
%%%% *************    Text stat'i     ********************************
%%%% *****************************************************************
\newpage
\pagenumbering{arabic}

{\Large \bf On phase transitions during collisions

\vspace{4pt}
near the black holes horizon}

\vspace{11pt}
{\bf
Andrey A. Grib${}^{1,2}$ and Yuri V. Pavlov${}^{3,4,*}$
}
\begin{abstract}
${}^{1}$ Theoretical Physics and Astronomy Department, The Herzen  University,
48~Moika, St.\,Petersburg, 191186, Russia;  andrei\_grib@mail.ru

${}^{2}$ A. Friedmann Laboratory for Theoretical Physics, St.\,Petersburg, Russia

${}^{3}$ Institute of Problems in Mechanical Engineering of
Russian Academy of Sciences,
61 Bolshoy, V.O., St.\,Petersburg, 199178, Russia

${}^{4}$ N.I.\,Lobachevsky Institute of Mathematics and Mechanics,
Kazan Federal University, Kazan, 420008 Russia

${}^{*}$ Correspondence: yuri.pavlov@mail.ru
\end{abstract}

    \begin{abstract}
\noindent
{\bf Abstract:}
    During particle collisions in the vicinity of the horizon of black holes,
it is possible to achieve energies and temperatures corresponding to
phase transitions in particle physics.
    It is shown that the sizes of the regions of the new phase are of
the order of the Compton length for the corresponding mass scale.
    The lifetime is also on the order of the Compton time.
    It is shown that inverse influence of the energy density in
the electro-weak phase transition in collisions on the space-time metric
can be neglected.

\vspace{7pt}
\end{abstract}

%%%% *****************************************************************
\section{\normalsize Introduction}
\label{secIn}

\hspace{\parindent}

    The works of A.A. Friedman~\cite{Friedmann22,Friedmann24}, written 100 years
ago, in which solutions were obtained for an expanding homogeneous isotropic
universe~\cite{Mostepanenko22},
are the theoretical basis of the modern standard cosmological model.
    The discovery in 1965 of relic radiation~\cite{PenziasWilson65,DickePeeblesRW65}
indicates that in the model of the expanding early Universe there were times
when the temperature of matter was so high that phase transitions predicted by
the theory of elementary particles could occur.
    There are three such phase transition in the standard model of particle
physics~\cite{KolbTurner,Linde,GorbunovRubakov}:

1) Between quark-gluon plasma and hadrons at energies $ E $ of the order of 200\,MeV.
    The corresponding temperature $ T = E/k_B\approx 10^{12}$\,K, where
$ k_B \approx 1.38 \cdot 10^{-23} $\,J/K is the Boltzmann constant,
may have taken place in the expanding Universe
during the order of $10^{-6} $\,s after the Big Bang.

2) Electro-weak phase transition at energies of the order of $ E_W \approx 100 $\,GeV.
    The corresponding temperature $ T_W\approx 10^{15}$\,K could take place
during the order of $10^{-12} $\,s after the Big Bang.

3) The grand unification phase transition at energies
$ E_{\rm GUT} \approx 10^{16}$\,GeV.
    The temperature corresponding to the energy of the grand unification phase
transition $ T_{\rm GUT} = E_{\rm GUT }/ k_B \approx 10^{29}$\,K,
may not have been achieved in the early universe,
in models with an inflationary stage in which the heating temperature
is significantly lower than~$ T_{\rm GUT} $.
    In models with a radiation-dominant stage in the early Universe,
the temperature~$ T_{\rm GUT} $ could be reached at times of the order
of $10^{-38} $\,s.

    The study of the properties of the matter at such temperatures and
the phenomena at these phase transitions is of undoubted theoretical interest.
    Is it possible to achieve such temperatures in experiments on the Earth?
    The maximum high temperature for macroscopic parts of the substance
is achieved at the time of nuclear explosion and can be on the order
of $10^8$\,K.
    It is significantly less than the temperature of even
the quark-gluon phase transition~\cite{Pasechnik}.

    The highest temperature achieved in experiments on Earth refers to
microscopic quantities of matter and is obtained when heavy element nuclei
collide in particle accelerators.
    Temperature in $4\cdot 10^{12}$\,K was obtained from a collision of
gold nuclei in Brookhaven National Laboratory (United States) in 2010~\cite{AdarePRL}.
    In 2012, it was reported that a temperature of $5 \cdot 10^{12}$\,K
was reached when the lead nuclei collided at the Large Hadron Collider~\cite{CNSPRL}.
    At such temperatures, hadron matter transforms to the quark-gluon
plasma state.
    However, such temperatures are more than two orders of magnitude less
than the temperature of the electro-weak phase transition.

    Thus, macroscopic amounts of matter in the state of phase transition
 of elementary particle physics in laboratories on Earth cannot be obtained,
 and microscopic amounts can be obtained only for the phase transition
 in the quark-gluon plasma state.

    Is it possible to observe matter at the temperatures of phase transitions
 of the particle physics in astrophysical processes at present?
    Brightly luminous accretion discs formed when matter falls into black holes
 have a visible temperature of hundreds of millions of Kelvin
 degrees~\cite{ShapiroTeukolsky}.
     As shown in our work~\cite{GribPavlov2022}, in the processes of collisions
of particles near the horizon of black holes, it is possible to achieve
energies in the system of the center of mass of colliding particles on
the order of the energy scale of the electro-weak phase transition.
    A summary of these results is presented in Section~\ref{secHEC}

    Here we will consider questions about the size of the regions of
the phase transition region obtained in a collision and the lifetime of such
a region.
    To do this, in Section~\hyperref[secPERCh]{3} we apply formulas for
the energy density and radiation intensity of a gas of relativistic particles.
    The possibility of obtaining an electro-weak phase transition in
a macroscopic volume during a collision in the vicinity of supermassive
black holes is studied in Section~\ref{PERChe}
    The influence of the matter energy-momentum tensor in the phase transition
region on the space-time metric will be evaluated in Section~\ref{sec5}

\vspace{4pt}
%%%% *****************************************************************
\section{\normalsize The high-energy collisions near the horizon of black holes}
\label{secHEC}

\hspace{\parindent}
    The Kerr metric of a rotating black hole~\cite{Kerr63} in the
Boyer-Lindquist coordinates~\cite{BoyerLindquist67} has the form
    \begin{equation}
d s^2 = \frac{\rho^2 \Delta}{\Sigma^2}\,c^2 d t^2 -
\frac{\sin^2\! \theta}{\rho^2} \Sigma^2 \, ( d \varphi - \omega d t)^2
\label{Kerr}
- \frac{ \rho^2}{\Delta}\, d r^2 - \rho^2 d \theta^2 ,
\end{equation}
    where
    \begin{equation} \label{Delta}
\rho^2 = r^2 + \frac{a^2}{c^2} \cos^2 \! \theta, \ \ \ \ \
\Delta = r^2 - \frac{2 G M r}{c^2} + \frac{a^2}{c^2},
\end{equation}
    \begin{equation} \label{Sigma}
\Sigma^2 = \left( r^2 + \frac{a^2}{c^2} \right)^2 -
\frac{a^2}{c^2} \sin^2\! \theta\, \Delta , \ \ \ \
\omega = \frac{2 G M r a}{\Sigma^2 c^2} ,
\end{equation}
   $G$ is the gravitational constant, $c$ is the speed of light,
$M$ and $ aM $ are the mass and angular momentum of the black hole,
respectively.
    We accept that  ${0 \le a \le G M/c }$.
    The event horizon of the Kerr black hole has the radial coordinate
    \begin{equation}
r = r_H \equiv \frac{G}{c^2} \left( M + \sqrt{M^2 - \left( \frac{a c}{G}
\right)^2} \right) .
\label{Hor}
\end{equation}

    According to~\cite{BSW} the squared energy of collision of two particles
with the mass~$m$ with the angular momenta $L_1$ and $L_2$ in
the center-of-mass system, which are nonrelativistic at infinity and are
freely incident on a black hole with the angular momentum~$aM$, is given
by the expression
    \begin{eqnarray}     \label{BSW}
\frac{E_{\rm c.m}^2}{m^2 c^4} = \frac{2}{x(x^2-2 x+A^2)} \Biggl[
2 A^2 (1+ x) - 2 A (l_1+l_2) - l_1 l_2 (x-2) + 2(x-1) x^2 \ \ \\
-\, \sqrt{2(A\!-\!l_2)^2 \!- l_2^2 x + 2 x^2 } \sqrt{2(A \! -\! l_1)^2
\!- l_1^2 x + 2 x^2 }
\Biggr], \nonumber
\end{eqnarray}
    where $x= r c^2/ G M$, $l= L c/ G m M$, and $A= a c/ G M$.
    The expression~(\ref{BSW}) has a singularity on the event horizon.
    In general case the limit value of the collision energy for two particle
with masses $m_1$, $m_2$, energies $E_1$, $E_2$ and angular momenta $J_1$, $J_2$ is
    \begin{equation}     \label{BSW1}
E_{\rm c.m}^2(r \to r_H) = \frac{c^6 ( J_{1H} J_2 \!- J_{2 H} J_1 )^2}{G^2 M^2
( J_{1H} \!-\! J_1) (J_{2H} \!-\! J_2) } +
m_1^2 c^4 \! \left[ 1 + \frac{ J_{2H} \!-\! J_2 }{ J_{1H} \!-\! J_1 } \right] +
m_2^2 c^4 \! \left[ 1 + \frac{ J_{1H} \!-\! J_1 }{ J_{2H} \!-\! J_2 } \right]\!,
\end{equation}
    where $ J_{n H} = 2 E_n r_H / A $.
    If the angular momentum of one of the particles tends to $ J_{n H} $,
then the expression for the energy~(\ref{BSW1}) diverges.
    This is the so-called Banados-Silk-West effect.
    Note that despite the unlimited increase in collision energy in
the center of mass system, the energy that can be extracted at a large distance
from a black hole cannot exceed $E_1 + E_2$
(assuming no Penrose effect~\cite{Penrose69}).
    This follows from the law of energy conservation.

    A particle having a critical angular momentum value can get from infinity
to the event horizon of a black hole only in the case of an extremely
rotating black hole $A=1$.
    In other cases, particles with large angular momentum values
are prevented from falling onto the horizon by the potential barrier of
the effective potential.
    As shown in~\cite{GribPavlov2010,GribPavlov2011}, the superhigh center-of-mass
energy can be achieved in multiple collisions near nonextreme black holes.
    To reach the horizon, particles incident from infinity should have an angular
momentum low in absolute value.
    The angular momentum of one of the particles necessary for a
high-energy collision can be acquired either in multiple collisions or in
the interaction with the electromagnetic field of the accretion disk.
    The similar effect for electrically charged black holes was discovered
in~\cite{Zaslavskii10b}.
    Real astrophysical black holes are surrounded by matter (for example,
they have an accretion disk).
    The possibility of particles colliding with unlimited energy near
the horizon of such ``dirty'' black holes also takes place~\cite{Zaslavskii10}.

    The value of the collision parameters corresponding to the temperature of
of the elementary particles phase transitions may depend on the type of black holes.
    In the case of Kerr black holes
    the estimates for the distance from the horizon, where the collision energies
required for phase transitions of elementary particles, can be achieved are given
in our work~\cite{GribPavlov2022}.
    So, for elementary particles with mass $m$ the value of temperature $T$ is reached
near the extreme rotating black hole at the distance
   \begin{equation}
r -r_H \approx 2 r_H \left( \frac{mc^2}{k_B T} \right)^2.
\label{R13dob}
\end{equation}
    For the proton mass the electro-weak temperature can be reached at a distance
$ r -r_H = 2 \times 10^{-4} r_H$.
    This amounts to tens of centimeters for stellar-mass black holes and hundreds
of thousands of kilometers for supermassive black holes.
    In the mechanism of multiple collisions near the horizon of (not extreme)
rotating black holes, such temperatures can be achieved at larger
distances~\cite{GribPavlov2022}.
    Therefore, collisions in which phase transition temperatures are reached can,
in principle, occur in the vicinity of stellar-mass black holes for elementary
particles, and in the case of supermassive black holes for macroscopic bodies.

    Next we estimate the size of the phase transition region and the lifetime
of the state with the new phase.

\vspace{4pt}
%%%% *****************************************************************
\section{\normalsize Size and lifetime of the new phase}
\label{secPERCh}

\hspace{\parindent}
    The energy density of the photon gas can be calculated by the known
formula: (see~(63,14) in~\cite{LL_V})
   \begin{equation}
\varepsilon = \frac{4 \sigma}{c} T^4,
\label{R13}
\end{equation}
    where $ \sigma $  is the Stefan-Boltzmann constant,
   \begin{equation}
\sigma = \frac{\pi^2 k_B^4}{60 \hbar^3 c^2} \approx
5.67 \cdot 10^{-8} \frac{\rm W}{\rm m^2\cdot K^4} ,
\label{R13SB}
\end{equation}
    $\hbar $ is the reduced Planck constant.
    At ultra-high temperatures, other elementary particles should also
contribute to the energy density of matter.
    Their contribution is taken into account using the factor $ g_{\rm eff} $
describing the number of effective massless degrees of freedom of particles
of the standard model of particle physics.
   \begin{equation}
\varepsilon =  g_{\rm eff} \frac{\pi^2 k_B^4}{30 \hbar^3 c^3} T^4 =
g_{\rm eff} \frac{2 \sigma}{c} T^4.
\label{R14}
\end{equation}
    Under such definition, the photon's contribution to $ g_{\rm eff} $
is 2, according to the photon's two polarization states.
    In general case one has~\cite{KolbTurner}
    \begin{equation}
g_{\rm eff} =  \sum_{i = \, {\rm bosons}} g_i \left( \frac{T_i}{T} \right)^4 +
\frac{7}{8} \sum_{i = \, {\rm fermions}} g_i \left( \frac{T_i}{T} \right)^4 .
\label{R15}
\end{equation}
    Here it is assumed that the equilibrium temperature $T_i$ of particles of
type $i$ may differ from $T$.
    For example, in the Universe at present the temperature of the cosmic
microwave background radiation is is equal to 2.7\,K, and estimates for
the temperature of the neutrino gas give 1.95\,K.
    Photon gas after the moment of the last collisions of cosmological neutrinos
with cosmological plasma at energies of 2--3 MeV was still heated up in
the process annihilation of cosmological positrons with electrons.

    The value of $g_{\rm eff} $ in standard model of particle physics depends
on temperature.
    For $T$ in the interval 1 MeV $< T< $ 100 MeV
taking into account neutrinos, electrons and positrons leads to
$g_{\rm eff} = 10.75$.
    At temperatures above 300 GeV, all standard model particles
(photon, $W^\pm$, $ Z^0 $ bosons, 8 gluons, 3 generations of quarks and leptons,
Higgs boson, must contribute to~(\ref{R14}), which
leads~\cite{KolbTurner} to the value $g_{\rm eff} = 106.75$.
    The graph of $g_{\rm eff} $ depending on temperature is presented
in~\cite{KolbTurner}, page~65, Fig.~3.5.

    Denoting $ k_B T = m c^2$, where $m$ is the characteristic mass scale,
we get from~(\ref{R14}) for the energy density of radiation of all types of
particles
   \begin{equation}
\varepsilon = g_{\rm eff}\frac{\pi^2 m^4 c^5}{30 \hbar^3 } =
g_{\rm eff} \frac{ \pi^2}{30}\, \frac{ m c^2}{ l_C^{\, 3} },
\label{R16}
\end{equation}
    where $ l_C= \hbar / m c $ is the (reduced) Compton wavelength
corresponding to the mass $m$.
    The pressure corresponds to a value three times less
       \begin{equation}
p =\frac{\varepsilon}{3} = g_{\rm eff} \frac{ \pi^2}{90}\,
\frac{ m c^2}{ l_C^{\, 3} }.
\label{R17}
\end{equation}

%%%%%%%%%%%%%%%%%%%%%%%%%%%%%%%%%%%%%%%%%%%%%%%%%%%%%%%%%%%
    The size $R_0$ of the area in which after a collision can form
the heated drop of a new phase of matter, we estimate from the relation
       \begin{equation}
E_{\rm c.m.} = \frac{4}{3} \pi R_0^3 \varepsilon ,
\label{ETsM1}
\end{equation}
    It is assumed that the region of the new phase is a sphere of radius $R_0$.
    Then one obtains
    \begin{equation}
R_0 = \frac{l_C}{\pi} \sqrt[3]{\frac{45}{2 g_{\rm eff}} \frac{E_{\rm c.m.} }{m c^2} }.
\label{ETsM2}
\end{equation}
    Assuming that the collision energy is of the order of magnitude
$ E_{\rm c.m.} \sim g_{\rm eff} m c^2 $, we find that the size of the region
phase transition is of the order of the Compton wavelength $l_C$ for a particle of mass $m$.

    Let us estimate the lifetime of a drop of a new phase formed as a result of a collision,
generalizing the formula for the radiation intensity of black body to the case
the presence of additional degrees of freedom described by the quantity $ g_{\rm eff} $
   \begin{equation}
J =  g_{\rm eff} \frac{\pi^2 k_B^4}{120 \hbar^3 c^2} T^4 =
\frac{g_{\rm eff}}{2} \sigma T^4.
\label{R14izl}
\end{equation}
    Let us write the energy balance equation for an infinitesimal time interval $d t$
    \begin{equation}
d(\varepsilon V ) = - J S d t,
\label{R14iz2}
\end{equation}
    where $V$ is the volume of new phase drop, $S$ --- its surface area.
    When obtaining estimates by order of magnitude, we will assume that
the drop is spherical, and its radius may depend on time due to expansion into
the surrounding space.
    We also assume that during the life of a drop of a new phase, thermodynamic
equilibrium takes place in it and, therefore, we can talk about the temperature of
the entire drop, the dependence of temperature on time and use formulas for
the equilibrium state of the corresponding relativistic gas.
    Then from~(\ref{R14iz2}) we get
    \begin{equation}
\frac{R}{3} d \varepsilon = - \left( J d t + \varepsilon d R \right) .
\label{R14iz3}
\end{equation}
    Using~(\ref{R14}) and~(\ref{R14izl}), we obtain the equation
    \begin{equation}
\frac{16}{3} \frac{R}{c} \frac{d T}{T} = - \left( 1 +
\frac{4}{c}  \frac{d R}{ d t} \right) d t .
\label{R14iz4}
\end{equation}
    By integrating this equation we get
    \begin{equation}
T(t) = T (t_0) \exp \left[
- \frac{3 c}{16} \int \limits_{t_0}^t \left( 1+ \frac{4}{c} \frac{d R}{ d t}
\right) \frac{d t}{R} \right].
\label{RTint}
\end{equation}
    If the drop radius does not change $ R \approx R_0 = {\rm const} $,
then the solution is
    \begin{equation}
T(t) = T (t_0) \exp \left[
- \frac{3 }{16} \frac{c(t-t_0)}{R_0} \right].
\label{RTint0}
\end{equation}
    Thus, the temperature decreases exponentially and the lifetime
of the new phase is of order $\tau \approx R_0/ c $.
    Since according to the equation~(\ref{ETsM2}), the size of the new phase region
is assumed to be Compton, then the lifetime corresponds to Compton time
$\tau_C =\hbar/( m c^2) $ for a particle of mass~$ m $, corresponding to
the phase transition energy.

    Taking into account the possible expansion of the area of the new phase,
the lifetime of the new phase can only decrease.
    Let's give formulas under the assumption of a constant expansion rate
$dR / d t = v \approx {\rm const}$.
    Then
    \begin{equation}
R(t) = R_0 + v ( t - t_0)
\label{R14iz5}
\end{equation}
    and after the integration of~(\ref{R14iz4}) we obtain a dependence of
the temperature of the region with a new phase on time
   \begin{equation}
T(t) = T(t_0) \left( 1 + \frac{v(t-t_0) }{R_0} \right)^{-\frac{3}{16}
\left( 4 + \frac{c}{v}\right)}.
\label{R14iz6}
\end{equation}
    In the limit $ v/c\to 0 $, one obtains the expression~(\ref{RTint0}).

    Thus, the lifetime of the new phase obtained in a collision of elementary
particles has the order of Compton time $\hbar/( m c^2) $ for a particle
of characteristic mass scale $ m $.
    For the quark-gluon phase transition this time is
$\tau \approx 3 \cdot 10^{-24}$\,s, for the electro-weak phase transition
$\tau \approx 7 \cdot 10^{-27}$\,s.

\vspace{4pt}
%%%% *****************************************************************
\section{\normalsize Phase transition in macroscopic volume}
\label{PERChe}

\hspace{\parindent}
    To perform a quark-gluon or electro-weak phase transition in
a macroscopic volume, it is necessary to collide with ultra-relativistic
energies of macroscopic amounts of matter.
    When ordinary macroscopic bodies collide with such energies, the regions of
the new phase can make up a macroscopic volume only if the density of
the bodies is comparable to the Compton density characteristic of
the phase transition of mass $ m $ (see~(\ref{R16})).
    Only in this case, the lifetime of the new phase can significantly exceed
the Compton time $\tau_C $.
    Such density of matter occurs only in neutron stars.
    Collisions of macroscopic objects with ultra-relativistic velocities
are possible in the vicinity of the horizon of extremal rotating
black holes~\cite{GribPavlov2022}.
    The collision of compact objects with star masses near supermassive
black holes was considered in~\cite{HaradaKimura11b}.

    When falling towards the event horizon of a black hole, macroscopic bodies can be
destroyed by tidal gravitational forces.
    Let's estimate the mass of black holes in which it is possible to fall
to the event horizon of neutron stars without destruction by tidal forces.
    For evaluation, we assume that a star is destroyed if the tidal forces for
the points of the center of mass and the surface exceed the force of attraction
of the points of the surface to the center of the falling body.
    Let us assume that the falling object (neutron star) is a uniform ball of
density $\rho$ and radius $R$.
    Also let's consider only the nonrotating black hole and radial tidal forces.
    Then the condition for the falling to the horizon without destruction has the form
   \begin{equation}
\frac{2 G M}{r_g^3} R < \frac{G 4 \pi \rho R^3}{3 R^2}
\label{razruh}
\end{equation}
    or (after simple transformations)
    \begin{equation}
M > \frac{c^3}{4 G^{3/2}} \sqrt{\frac{3}{\pi \rho }}, \ \ \ \ \
\frac{M}{M_\odot} > 1.9 \cdot 10^8 \sqrt{\frac{\rho_w}{\rho}},
\label{razruh2}
\end{equation}
    where $M_\odot$ is the Sun mass,
$\rho_w = 10^3$\,kg/m$^3$ is the water density.
    Neutron stars have the density $\rho \sim 10^{17} - 10^{18}$\,kg/m$^3$.
    Therefore, neutron stars fall to the horizon of black holes
with mass $M > 20 M_\odot $ without destruction.
    Of course, a collision with ultra-relativistic velocities of neutron stars
in the vicinity of a massive black hole should be considered a very unlikely event.
    Estimates in~\cite{GribPavlov2022} show that in the collision near
the vicinity of the horizon extreme rotating black hole with a mass of $10^9 M_\odot$
in points with radial coordinate $r_H + 7\cdot 10^5$\,km
the maximum collision energy in the center system mass can reach $100mc^2$.
    In nucleon-nucleon collisions, this is the electro-weak unification energy.
    The masses of neutron stars range from one to three solar masses,
and their radii are about 10 --- 20 km.
    The gravitational radius of a black hole with a mass of 100 solar
masses is approximately 300 km.
    Therefore, with such collision energy of two neutron stars,
a black hole should form and it will not be possible to obtain a substance in
a state of an electro-weak phase transition outside the event horizon.

     Thus, it is impossible to obtain macroscopic quantities of a substance
with electro-weak phase transition with a lifetime significantly exceeding
Compton time for the electro-weak scale due to collisions in
the vicinity of black holes.

\vspace{4pt}
%%%% *****************************************************************
\section{\normalsize The influence of spontaneous symmetry breaking on
the space-time metric}
\label{sec5}

\hspace{\parindent}
    Let us consider a real scalar field with self-action~\cite{Okun}
    \begin{equation}
V(\varphi) = - \frac{\mu^2}{2} \varphi^2 + \frac{\lambda^2}{4} \varphi^4
+ \frac{\mu^2}{4\lambda^2}.
\label{gp3}
\end{equation}
    Here, $\mu = \tilde{\mu} c/\hbar$,
$\tilde{\mu}$ is a mass parameter,
$\lambda $ is the dimensionless self-action constant.
    Stable equilibrium states of such field are located at two points
    \begin{equation}
\varphi = \pm \varphi_0, \ \ \ \ \varphi_0 = \frac{\mu}{\lambda} .
\label{gp4}
\end{equation}
    The potential function~(\ref{gp3}) can be written as
    \begin{equation}
V(\varphi) = \frac{\lambda^2}{4} \left( \varphi^2 - \varphi_0^2 \right)^2 .
\label{gp3d}
\end{equation}
    Both lower states have zero energy, and the unstable equilibrium
with $\varphi=0$ has an energy density
    \begin{equation}
\varepsilon = \hbar c V(0) = \hbar c \frac{\mu^4 }{4 \lambda^2} .
\label{gp5}
\end{equation}
    Using the representation $\varphi=\varphi_0 + \chi$ one obtains
    \begin{equation}
V(\chi) = \lambda^2 \varphi_0^2 \chi^2 + \lambda^2 \varphi_0 \chi^3 +
\frac{\lambda^2}{4} \chi^4.
\label{R9}
\end{equation}
    Thus, the mass of the $\chi $ field is
$ \sqrt{2} \lambda \varphi_0 = \sqrt{2} \mu $.
    In the case of Higgs boson $m_H = 125.3$\,GeV and we have
    \begin{equation}
\varepsilon_H = \hbar c  V (0) = \hbar c \, \frac{m_H^4 c^4}{16 \hbar^4 \lambda^2}
= \frac{1}{16 \lambda^2}\, \frac{m_H c^2}{(l_{C}^{H})^3},
\label{gp05}
\end{equation}
    where $ l_{C}^H = \hbar / (m_H c) $.

    For the electro-weak interaction, quantum corrections lead to
the limitation~\cite{Okun}
    \begin{equation}
\lambda \ge \alpha = \frac{e^2}{4 \pi \varepsilon_0 \hbar c} \approx \frac{1}{137} ,
\label{R8}
\end{equation}
    where $e$ is the elementary electric charge, $\varepsilon_0 $ is the electric constant.

    To estimate the inverse influence of the scalar field on the curvature of
space-time we use Einstein's equations
     \begin{equation}
R_{ik} -\frac{1}{2} R g_{ik} + \Lambda g_{ik} = - 8 \pi \frac{G}{c^4} (T_{ik}^{(0)}
+ T_{ik}),
\label{Eur}
\end{equation}
    where $\Lambda$ is the cosmological constant,
$T_{ik}^{(0)}$ is the energy-momentum tensor of background matter.
    The energy-momentum tensor for a constant scalar field with minimal coupling
with curvature is~\cite{GMM}
     \begin{equation}
T_{ik} = g_{ik} \hbar c V(\varphi)
\label{Tikcs}
\end{equation}
    and is similar to the contribution of an additional cosmological constant.
    On the appearance of a non-zero cosmological constant under spontaneous
symmetry breaking, it was indicated in the work~\cite{Grib67}.
    Phase transition in electro-weak interactions was discussed in cosmology
by Kirzhnits and Linde~\cite{KirzhnitsLinde74,KirzhnitsLinde76},
Weinberg~\cite{Weinberg74} and others.
    Estimates of changes in the value of the cosmological constant during phase
transitions in the early Universe were made at work~\cite{Linde74}.

    If there is only a constant scalar field and the energy-momentum tensor of
the background matter is equal to zero $ T_{ik}^{(0)} = 0 $,
then the solution to the Einstein equations~(\ref{Eur}) will be the de Sitter
space-time.
    In de Sitter space one has
    \begin{equation}
R_{ik} = \frac{R}{4} g_{ik},
\label{RSg}
\end{equation}
    and it follows from~(\ref{Eur})
    \begin{equation}
R = 4 \left( \Lambda +  l_{\rm Pl}^2\, 8 \pi V(\varphi) \right),
\label{RSg2}
\end{equation}
    where    $ l_{\rm Pl} $ is the Planck length
    \begin{equation}
l_{\rm Pl}= \sqrt{ \frac{G \hbar }{c^3}} = 1.6162 \cdot 10^{-35}\,{\rm m}.
\label{Plle}
\end{equation}
    For the electro-weak case under $\varphi=0$ from~(\ref{gp05}) we have
    \begin{equation}
R = 4 \left( \Lambda +  \frac{\pi}{2 \lambda^2} \frac{l_{\rm Pl}^2}{ (l_{C}^H)^4} \right).
\label{RSg3}
\end{equation}
    For the radius of curvature we obtain
    \begin{equation}
r \sim \frac{(l_{C}^H)^2}{l_{\rm Pl}} \sim 0.1\,{\rm m}.
\label{RSg4}
\end{equation}
    This value is many orders of magnitude greater than the Compton wavelength
of the particle and the size of the region in which the phase transition occurs.
    It should be expected that in order for special collisions with ultra-high
energy to occur, in such volumes $r^3$ there must be a large number of particles
falling onto the black hole.
    Their total mass will be much greater than the mass of the electro-weak scale.
    Thus, the inverse effect of energy density in the electro-weak phase transition
 in collisions on the space-time metric can be neglected.

\vspace{4mm}
%%%% *****************************************************************
\section{\normalsize Conclusion}
\label{secConcl}

\hspace{\parindent}
     An integral part of the standard model of particle physics is
the mechanism of spontaneous symmetry breaking.
     The discovery at the Large Hadron Collider of the Higgs boson in 2012 makes
us take seriously the possibility of a phase transition from one vacuum
to another at high temperatures, as it is the case in quantum
non-relativistic many-body theory, where the ground state plays the role
of the vacuum.
    In our work~\cite{GribPavlov2022} was shown that in the processes
of collisions  of particles near the horizon of black holes,
it is possible to achieve energies in the system of the center of mass
of the order of the energy scale of the electro-weak phase transition.

    In this article we showed that the region of phase transition in such
collisions is microscopic.
    In order of magnitude, the size of the region is equal to the Compton
wavelength of the Higgs boson.
    Using formulas for black body radiation, we show that the lifetime
of such region is of the order of the Compton time for
the electro-weak phase transition scale.

    During a phase transition in the case of spontaneous symmetry breaking,
the energy-momentum tensor corresponds to the emergence of an effective
cosmological constant.
  It is shown that for phase transitions during particle collisions
its influence on the space metric in the phase transition region
can be neglected.

    Note that despite the short time existence and microscopic volumes of
a new phase of matter during an electro-weak phase transition in collisions in
the vicinity of the black hole horizon, its very existence is of fundamental
importance for the study of elementary particle physics in
the ultra-high energy region, unattainable on Earth.

\vspace{7pt}
\noindent
{\bf Author Contributions:}
A.A.G. and Y.V.P. have contributed equally to all parts of this work.
All authors read and agreed to the published version of the manuscript.

\vspace{7pt}
\noindent
{\bf Funding:}
This research received no external funding

\vspace{7pt}
\noindent
{\bf Data Availability Statement:}
Data are contained within the article.

\vspace{7pt}
\noindent
{\bf Conflicts of Interest:} The author declares no conflict of interest.

%%%% *****************************************************************


\begin{thebibliography}{99}

\bibitem{Friedmann22}
Friedmann, A.A. \"{U}ber die Kr\"{u}mmung des Raumes.
{\em Z. Phys.} {\bf 1922}, {\em 10}, 377--386.
[\href{https://doi.org/10.1007/BF01332580}{CrossRef}]

\bibitem{Friedmann24}
Friedmann, A.A. \"{U}ber die M\"{o}glichkeit einer Welt mit konstanter
negativer Kr\"{u}mmung des Raumes.
{\em Z. Phys.} {\bf 1924}, {\it 21}, 326--332.
[\href{https://doi.org/10.1007/BF01328280}{CrossRef}]

\bibitem{Mostepanenko22}
Klimchitskaya, G.L.; Mostepanenko, V.M.
Centenary of Alexander Friedmann's prediction of the Universe expansion and
the quantum vacuum.
{\em Physics} {\bf 2022}, {\em 4}, 981--994.
[\href{https://doi.org/10.3390/physics4030065}{CrossRef}]

\bibitem{PenziasWilson65}
Penzias, A.A.; Wilson, R.W.
Excess antenna temperature at 4080 Mc/s.
{\em Astrophys. J.} {\bf 1965}, {\em 142}, 419--421.
[\href{https://doi.org/10.1086/148307}{CrossRef}]

\bibitem{DickePeeblesRW65}
Dicke, R.H.; Peebles, P.J.E.; Roll, P.G.; Wilkinson, D.T.
Cosmic black-body radiation.
{\em Astrophys. J.} {\bf 1965}, {\em 142}, 414--419.
[\href{https://doi.org/10.1086/148306}{CrossRef}]

\bibitem{KolbTurner}
Kolb, E.W.; Turner,  M.S. {\em The Early Universe};
Addison-Wesley: Redwood City, 1990.

\bibitem{Linde}
Linde, A. {\em Particle Physics and Inflationary Cosmology};
Harwood Academic Publication: New York, 1990.

\bibitem{GorbunovRubakov}
Gorbunov, D.S.; Rubakov, V.A. {\em Introduction to the Theory of the Early
Universe. Hot Big Bang Theory};
World Scientific: Singapore, 2018.

\bibitem{Pasechnik}
Pasechnik, R.; $\check{S}$umbera, M.
Phenomenological review on quark-gluon plasma: concepts vs. observations.
{\em Universe} {\bf 2017}, {\em 3}, 7.
[\href{https://doi.org/10.3390/universe3010007}{CrossRef}]

\bibitem{AdarePRL}
Adare, A. et. al. [PHENIX Collaboration].
Enhanced production of direct photons in Au + Au collisions at
$ \sqrt{s_{NN}} = 200 $ Gev and implications for the initial temperature.
{\em Phys. Rev. Lett.} {\bf 2010}, {\em 104}, 132301.
[\href{https://doi.org/10.1103/PhysRevLett.104.132301}{CrossRef}]

\bibitem{CNSPRL}
Chatrchyan S. et al. [CMS Collaboration].
Measurement of the pseudorapidity and centrality dependence of the transverse
energy density in Pb-Pb collisions at $ \sqrt{s_{NN}} = 2.76 $ Tev.
{\em Phys. Rev. Lett.} {\bf 2012}, {\em 109}, 152303.
[\href{https://doi.org/10.1103/PhysRevLett.109.152303}{CrossRef}]

\bibitem{ShapiroTeukolsky}
Shapiro, S.L.; Teukolsky, S.A.
{\em Black Holes, White Dwarfs, and Neutron Stars. The Physics of Compact Objects};
Wiley Int. Publ.: New York, 1983.

\bibitem{GribPavlov2022}
Grib, A.A.; Pavlov, Yu.V. On phase transitions near black holes.
{\em JETP Lett.} {\bf 2022}, {\em 116}, 493--499.
[\href{https://doi.org/10.1134/S0021364022601907}{CrossRef}]

\bibitem{Kerr63}
Kerr, R.P.
{Gravitational field of a spinning mass as an example of algebraically
special metrics},
{\em Phys. Rev. Lett.} {\bf 1963}, {\em 11}, 237--238.
[\href{https://doi.org/10.1103/PhysRevLett.11.237}{CrossRef}]

\bibitem{BoyerLindquist67}
Boyer, R.H.; Lindquist, R.W.
{Maximal analytic extension of the Kerr metric},
{\em J. Math. Phys.} {\bf 1967}, {\em 8}, 265--281.
[\href{https://doi.org/10.1063/1.1705193}{CrossRef}]

\bibitem{BSW}
Banados, M.; Silk, J.; West, S.M.
{Kerr black holes as particle accelerators to arbitrarily high energy},
{\em Phys. Rev. Lett.} {\bf 2009}, {\em 103}, 111102.
[\href{https://doi.org/10.1103/PhysRevLett.103.111102}{CrossRef}]

\bibitem{Penrose69}
Penrose, R.;
Gravitational collapse: the role of general relativity,
{\em Rivista Nuovo Cimento} {\bf 1969}, {\em I} Num. Spec., 252--276.
[\href{http://adsabs.harvard.edu/abs/1969NCimR...1..252P}{Ref}]

\bibitem{GribPavlov2010}
Grib, A.A.; Pavlov, Yu.V.
{On the collisions between particles in the vicinity of rotating black holes},
{\em JETP Lett.} {\bf 2010}, {\em 92}, 125--129.
[\href{https://doi.org/10.1134/S0021364010150014}{CrossRef}]

\bibitem{GribPavlov2011}
Grib, A.A.; Pavlov, Yu.V.
{On particle collisions in the gravitational field of the Kerr black hole},
{\em Astropart. Phys.} {\bf 2011}, {\em 34}, 581--586.
[\href{https://doi.org/10.1016/j.astropartphys.2010.12.005}{CrossRef}]

\bibitem{Zaslavskii10b}
Zaslavskii, O.B.
Acceleration of particles by nonrotating charged black holes.
{\em JETP Lett.} {\bf 2010}, {\em 92}, 571--574.
[\href{https://doi.org/10.1134/S0021364010210010}{CrossRef}]

\bibitem{Zaslavskii10}
Zaslavskii, O.B.
Acceleration of particles as a universal property of rotating black
holes.
{\em Phys. Rev. D} {\bf 2010},  {\em 82}, 083004.
[\href{https://doi.org/10.1103/PhysRevD.82.083004}{CrossRef}]

\bibitem{LL_V}
Landay, L.D.; Lifshitz, E.M. {\em Statistical Physics. Part~1};
Pergamon Press: Oxford, 1980.

\bibitem{HaradaKimura11b}
Harada, T.; Kimura, M.
Collision of an object in the transition from adiabatic inspiral
to plunge around a Kerr black hole.
{\em Phys. Rev. D} {\bf 2011}, {\em 84}, 124032.
[\href{https://doi.org/10.1103/PhysRevD.84.124032}{CrossRef}]

\bibitem{Okun}
Okun, L.B. {\em Leptons and quarks}; North-Holland: Amsterdam, 1985.

\bibitem{GMM}
Grib, A.A.; Mamayev, S.G.; Mostepanenko, V.M.
{\it Vacuum Quantum Effects in Strong Fields};
Friedmann Lab. Publ.: St.\,Petersburg, 1994.

\bibitem{Grib67}
Grib, A.A. $CP$-noninvariance in $K^0$-meson decays and
nonequivalent representations in quantum field theory.
{\em Vestnik LGU} {\bf 1967}, {\em 22}, Vyp.~4, 50--56.

\bibitem{KirzhnitsLinde74}
Kirzhnits, D.A.; Linde, A.D. Relativistic phase transitions.
{\em Sov. Phys. JETP.} {\bf 1975}, {\em 40}, 628.

\bibitem{KirzhnitsLinde76}
Kirzhnits, D.A.; Linde, A.D.
Symmetry behavior in gauge theories.
{\em Annals of Physics} {\bf 1976}, {\em 101}, 195--238.
[\href{https://doi.org/10.1016/0003-4916(76)90279-7}{CrossRef}]

\bibitem{Weinberg74}
Weinberg, S. Gauge and global symmetries at high temperature.
{Phys. Rev.~D {\bf 1974}, {\em 9}, 3357--3378}.
[\href{https://doi.org/10.1103/PhysRevD.9.3357}{CrossRef}]

\bibitem{Linde74}
Linde, A.D. Is the cosmological constant a constant?
{\em JETP Lett.} {\bf 1974}, {\bf 19}, 183--184.

\end{thebibliography}
\end{document}